\documentclass[prb,nobibnotes,floatfix,superscriptaddress]{revtex4}

\usepackage{amsfonts,amsmath,graphicx,epsfig,latexsym}
\usepackage{subfigure}

\begin{document}

\title{Mechanism behind the switching of current induced by a gate field in a semiconducting nanowire junction }

\author{Subhasish Mandal}
\affiliation
{Department of Physics, Michigan Technological University, Houghton MI 49931}

\author{Ranjit Pati}
\affiliation
{Department of Physics, Michigan Technological University, Houghton MI 49931}

\begin{abstract}

{\footnotesize
We propose a new orbital controlled model to explain the gate field induced switching of current in a semiconducting PbS-nanowire junction. 
A single particle scattering formalism in conjunction with a {\it posteriori} density functional approach involving hybrid functional 
is used to study the electronic current; both first and higher order Stark effects are explicitly treated in our model.
Our calculation reveals that after a threshold gate-voltage, orbital mixing produces {\it p}-components at the S atoms in the participating orbitals. 
This results in an {\it inter-layer} 
orbital interaction that {\it allows electron to delocalize along the channel axis}. As a consequence a higher conductance state is found.
A similar feature is also found in a PbSe nanowire junction, which suggests that this model can be used universally 
to explain the gate field induced switching of current in lead-chalcogenide nanowire junctions.

}
\end{abstract}

\maketitle
\newpage
\section{Introduction}

While the size of the conventional silicon based field effect transistor is inching toward its fundamental limit of miniaturization, the quantum 
controlled, semi-conducting nano-wire has emerged as one of the promising options to meet the physical challenges 
imposed by quantum mechanics\cite{lieber1,circuit}.
Field effect transistors (FET), whose main function is to switch {\it source-drain} current upon the application of {\it gate} field, have been 
fabricated from semiconducting nano-wires (NW) and nano-crystal arrays\cite{lieber1,circuit,ScienceT}; 
the switching speed for such devices, in some instances, have been found to 
surpass that of the conventional semiconductor-FET. Particularly, IV-VI\cite{Talapin1,Talapin2,Talapin3,ScienceT} semi conducting nano-wires 
based FET devices have been shown to 
have huge conductivity and high current gain, which are {\it key requisites} for an integrated circuit. The controlled synthesis 
of these NWs with diameter ranging from 1.2 nm to 10 nm have been reported\cite{pbsnw1,pbsnw2,pbsnw3}.
  
  Despite the rapid development on the experimental side, which provides an ample opportunity for theorist to test their models, only nominal 
theoretical efforts are made to understand the {\it quantum} {\it phenomenon} that dictates current modulation in 
such a nanowire junction (particularly NW of few nm dimension); thus far, 
no unswerving 
interpretation exists. Since the dimension of the 
{\it channel} is in the nanoscale regime, the electronic property and/or the field effect switching mechanism is expected to be different than that of 
the conventional FET. 
In this letter, we present a first principles quantum transport study in a strongly coupled, single PbS nanowire (PbSNW) junction (Fig. 1) to unravel the
mechanism responsible for the gate field induced switching of current.
 We have used the same gold electrode to form the nanowire-{\it 
lead} junction as used in the experiment\cite{Talapin1, Talapin2, Talapin3,ACS}. Particularly, we try to answer several fundamental questions: 
How does the gate field affect the intrinsic electronic 
structure of the nanowire? Can we control the number of participating orbitals of the NW-{\it channel} via gate field? Which are the orbitals that contribute 
to the conductance? Can we manipulate the shape of the orbital via gating? 
and finally, is there a universal model that would explain the observed gate field induced switching not  
only in PbSNW but also in other lead-chalcogenide nanowire?

A single particle scattering formalism in conjunction with a density functional
 approach is used to study the electronic current\cite{Max,dutta,Ratner,Guo1,Guo2,Sanvito,Pati}.
We have included self-consistently both first and higher order Stark effects in our model. 
Our calculation shows, upon application of transverse gate field, the symmetry of the wavefunction is broken along the direction perpendicular 
to the channel axis; the participating molecular orbitals (MO) start to localize in the direction of gate field resulting in a shifting of 
unoccupied energy levels towards the Fermi energy. After a threshold gate voltage of -3.7 V, orbital mixing produces a dominant {\it p}-component at the 
S-atom in the participating MOs. This results in an {\it inter-layer} orbital interaction leading to electron delocalization along the channel axis.
 As a consequence a higher conductance state is found. The higher conductance state is referred here as the ON state and the lower conductance state
 prior to the threshold value (-3.7 V) is termed as the OFF state. It should be noted that in the OFF state, S atom in the participating orbital 
has only {\it s}-component. A similar feature has been observed in a PbSe nanowire junction, where {\it s} components are found at the Se atom in the OFF
state. Thus this orbital controlled model can be used universally to understand the observed gate field induced switching behavior in  
lead-chalcogenide NW junction.
 
The rest of the paper is organized as follows. The modeling of the device is described briefly in Section II followed by Results
and Discussions in Section III.  Our main findings are summarized in Section IV.

\section{ Modeling the Device}
 
   	For our calculations, we have used a real space approach in which the single determinant wave function is constructed from a 
finite set of Gaussian atomic orbitals\cite{Gaussian}. 
This allows us to partition the {\it open} NW device structure (Fig. 1) into three parts; 
the first part is the scattering region comprised of a finite NW of length $\sim $ 1.2 nm and diameter $\sim$ 1.17 nm , 
the second part is part of the {\it lead} that is strongly coupled to the NW and is represented only by a finite number of gold atoms
 (five gold atoms on each side), and the third part is the unperturbed electrode part which is assumed to retain the bulk behavior of gold. 
The atomic level structural details for the finite part of the PbSNW is taken from the optimized structure of an infinite NW,
 grown in the observed [100] direction. The later structure having six Pb and six S atoms in each layer along the growth direction 
with a lattice parameter of 6 $\AA$ was calculated using the periodic DFT\cite{Subhasish}. 
Only a five layer NW-structure along the growth direction is considered to build the symmetric 
junction with the {\it lead} (Fig. 1). The atomic composition of the {\it lead} is taken from the Au [100] surface to avoid the lattice mismatch at the 
NW-{\it lead} interface. To realize a {\it strongly} coupled junction between the NW and the {\it lead}, the interfacial distance is varied to determine the 
optimum distance (2.80 $\AA$) where the repulsive interaction is minimum. 

 Electron transport is a nonequilibrium (NEB) process\cite{Max,dutta} that requires calculation of the electronic structure of the device (Fig.1) under 
applied bias. The NEB situation refers to the bias condition when the self consistent (SC) potential at the {\it lead} 
on one side 
($V_\mathcal{L}$) is different from that on the opposite side ($V_\mathcal{R}$); the equilibrium (EB) situation is described by $V_\mathcal{L}=V_\mathcal{R}$. To replicate the NEB situation in our symmetric NW-junction, an electric-dipole interaction term is included in 
the Hamiltonian of the {\it active} region (NW+ finite {\it lead}) of the device as:
$\mathcal{H}(\vec{\varepsilon_{d}})= \mathcal{H}(0)+\vec{\varepsilon_{d}} \cdot \sum_{i}\vec{r}_i  $,  where  
$\mathcal{H}(0)$ is the Hamiltonian in the absence of electric field;
$\vec{\varepsilon_{d}}$ is the applied dipole electric field along the axis parallel to the direction of current flow (z-axis), 
and $\vec{r_i}$ is the coordinate of the $i^{th}$ electron; charging effect on the NW is considered by including a finite part of the lead.
The self-consistent inclusion of dipole interaction term permits us to include both first and higher order Stark effects, which is also evident 
from the comparison of total energy in the {\it active} region for different $\vec{\varepsilon_d}$; 
a non-linear change in energy by increasing the strength of the  $\vec{\varepsilon_d}$
 confirms the inclusion of higher order effects. This approach allows us to create an imbalance in charge carrier between the two {\it leads} as a 
function of the dipole field strength($\varepsilon_d$); on one {\it lead} there is a charge surplus ({\it source}) and on the other {\it lead} there is a 
charge depletion ({\it drain}) resulting local dipoles often referred to as {\it residual resistivity dipoles}\cite{Max}.  
This intrinsic charge imbalance between the two leads is also reflected from the potential profile summarized in Fig. 2. 
The relative electrostatic potential (REP) in Fig. 2 is calculated by subtracting the average potential at each atomic site in a layer at the 
EB condition from that at the NEB condition. A linear drop in the REP value along the wire axis is noted. The magnitude of the potential drop at 
both the junctions are equal confirming the NW-junction to be symmetric. A non linear change in the REP values with different $\vec{\varepsilon_d} $ 
elucidates the nonlinear response of the field. 
The REP values at the left and right Au {\it lead}, which are assumed to be at same potential with semi-infinite electrodes on left and right 
respectively, are used to calculate $V_\mathcal{L}$ and $V_\mathcal{R}$.
The electro chemical potentials at the two semi-infinite contacts are obtained as: 
$\mu_{\mathcal{L,R}}=V_\mathcal{L,R} \mp k_BT$\cite{Subhasish2,PPP}. 
A small thermal smearing term ($\sim k_BT$) in $\mu_{\mathcal{L}}$ and $\mu_{\mathcal{R}}$ takes into account the electronic temperature at the contact in the NEB condition;  the potential difference between source and drain ($V_{sd}$)  
is then obtained from the difference of $\mu_\mathcal{L}$ and $\mu_\mathcal{R}$. In order to simulate the effect of electrostatic gating, we have included an additional 
dipole interaction term ($\vec{\varepsilon_{g}} \cdot \sum_{i}\vec{r}(i)  $)in the Hamiltonian; the dipole field $\vec{\varepsilon_g}$ is applied along the direction perpendicular to the {\it channel} axis and is 
referred to as the transverse gate field in this article.
In our calculation, we have used a {\it posteriori} hybrid density functional method (B3LYP) that includes a portion of the exact Hartree-Fock exchange.
The LANL2DZ effective core potential basis set, which includes scalar relativistic effect, is used to describe the Pb and Au atom in 
the device; a triple zeta augmented by polarization function (6-311G*) basis set is used for the S.  
Subsequently, we recourse to  {\it implicit} {\it bias-dependent} Green's function approach\cite{Pati,PPP} to couple the finite NW to the infinite 
electrode via the finite self-energy 
functions; coherent, single particle scattering formalism is used to calculate the electronic current.  

\begin{figure}
\epsfig{figure=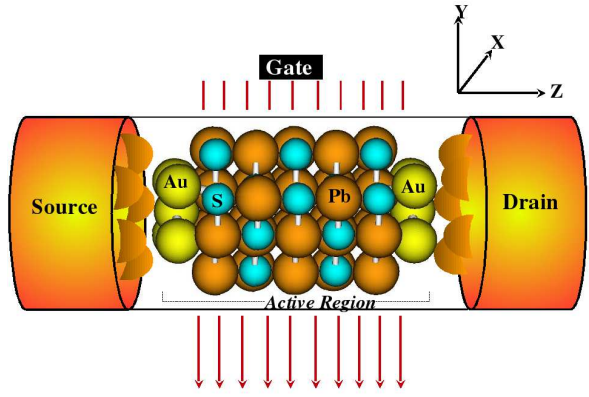, width=220pt}
\caption{
(Color online) Schematic representation of a PbS nanowire junction; solid arrows show the direction of the applied gate field. 
}

\end{figure}
\begin{figure}
\epsfig{figure=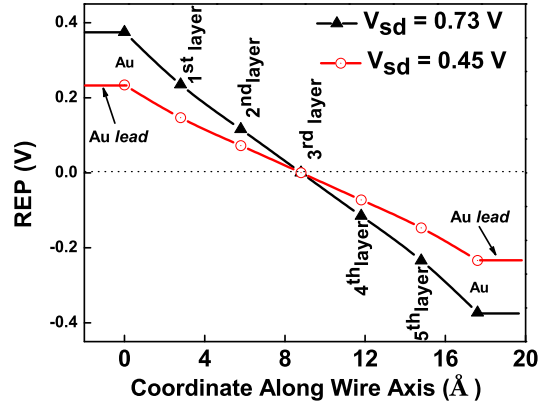, width=230pt}
\caption{
(Color online) 
Electrostatic potential profile of the NW junction in the absence of gate bias for two different $V_{sd}$.
}
\end {figure}

\section{Results and Discussions}
\subsection{Current-voltage Characteristics}

The calculated current-voltage ($I_{sd}$-$V_{sd}$) characteristic for PbSNW as a function of $\varepsilon_g$ is summarized in Fig. 3. The strength of 
$\varepsilon_g$ is mapped to the gate potential, $V_g$, by calculating the potential drop 
self-consistently between the terminal atomic layers of the NW along 
the direction of gate field. For $V_g$=0 V, a steady increase in current ($I_{sd}$) is noted with the increase of $V_{sd}$. Changing the
 $V_g$ from 0 V to -3.7 V, though an analogous linear increase in $I_{sd}$ as a function of $V_{sd}$ is observed, the magnitude of the  current is 
found to be higher ($\sim$ 1.5 times at $V_{sd} \sim$ 0.7 V) for $V_g$=-3.7 V. A further change of $V_g$ from -3.7 V to -5.6 V reveals 
a considerable increase in $I_{sd}$. Changing the $V_g$ from -5.6 V to -7.6 V, a non-linear feature in $I_{sd}$ is noted.  
The current at $V_{sd}$=0.74 is found to be 2.32 times higher for $V_g$=-5.6 V than that 
for $V_g$=-3.7 V; increasing the negative gate potential from -5.6 V to -7.6 V, 1.53 times higher current is found at $V_{sd} \sim $0.7 V.
 Thus comparing between $V_g$=0 V and $V_g$=-7.6 V, $\sim$ 5 times increase in $I_{sd}$ is found at $V_{sd}$ $\sim$0.7 V. 
To illustrate this behavior, we have plotted $I_{sd}$ as function of $V_g$ in the inset of Fig. 3; a fixed $V_{sd}$ 
is used. First a slow increase in current (OFF state) is noted till the value of $V_g$ reaches a threshold value ($V_{g}^{th}$)
of -3.7 V. After $V_g$=-3.7 V, 
a steep increase in current (ON state) is observed with the increase of gate potential resulting in a large change in the slope of $I_{sd}-V_g$.
The calculated ON/OFF current ratio value is found to be 
6.28 at $V_{sd}$ of 0.54 V  between  $V_g$=0 V and  $V_g$=-7.6 V. A similar switching feature is also found in a PbSe NW junction (Fig. 4).
It should be noted that the ON/OFF current ratio 
of $\sim$ 3.75 between $V_g$=0 V and $V_g$=-8 V at $V_{sd}$ of 0.5 V is observed in a recent experiment, where a 
single PbS nanowire is used as a {\it channel}.
The magnitude of $I_{sd}$ reported in the experiment is in the nA range in contrast to the $\sim \mu$A current observed in
our calculations.
Several reasons could be attributed to the observed differences in $I_{sd}$. First, in the experiment the {\it channel} length and diameter were $10^3$  and
150 nm respectively, where the diffusive transport could be the prevalent mechanism. In contrast, we have considered the {\it channel} length and diameter to be 1.2 nm and 1.17 nm for practical purposes.  
Considering an approximate exponential decay in current with the length ($l$) for the nanowire used in the experiment ($\sim e^{-\beta l};
 \beta$-decay constant), 
we would expect the measured current to be of the order of $\mu$A for a few nm channel length, which has also been reported experimentally in 
single PbSe semiconducting nanowire junction\cite{Talapin2}. In addition, we have considered an ideal, defect-free nanowire junction.
The magnitude of higher current observed in our calculation is also not surprising considering 
the use of static exchange and correlation potential instead of the true dynamical exchange correlation corrected potential\cite{evers,Runge,Max2,Max3}. 
However, the consistent increase of calculated current upon increasing negative gate bias as observed in the experiments\cite{Talapin2,ACS} reaffirms on the 
switching phenomenon replicated here. 

\begin{figure}
\epsfig{figure=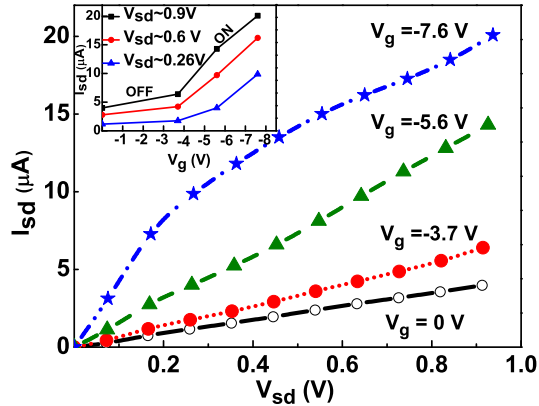,width=230pt}
\caption{ (Color online) Current-voltage characteristics with different $V_g$ for PbS nanowire junctions.
The insets show $I_{sd}$-$V_g$ plot for different $V_{sd}$.}
\end{figure}

\begin{figure}
\epsfig{figure=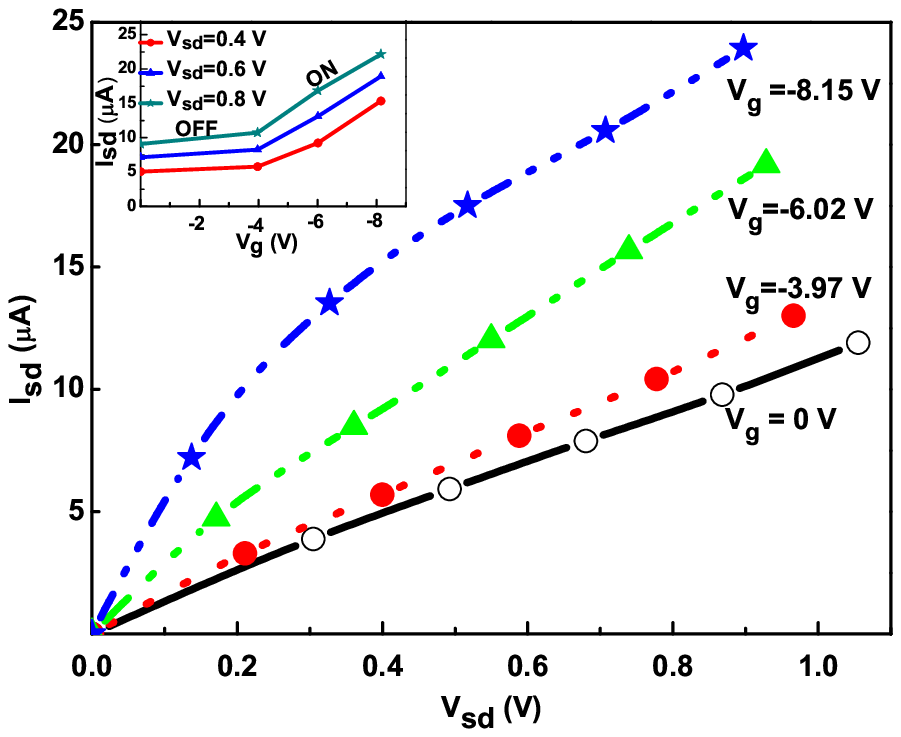,width=230pt}
\caption{ (Color online) Current-voltage characteristics with different $V_g$ for PbSe nanowire junctions.
The insets show $I_{sd}$-$V_g$ plot for different $V_{sd}$.}
\end{figure}

\begin{figure}
\epsfig{figure=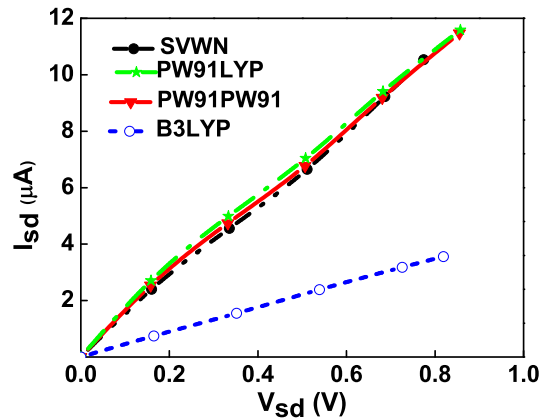,width=230pt}
\caption{ (Color online) Current-voltage characteristics of PbSNW for different exchange-correlation functionals at $V_g$=0 V. }
\end{figure}

 To examine, whether the increase in magnitude of current is due to the use of an implicit orbital 
dependent B3LYP functional approach, we have calculated the current in the same device geometry using different exchange-correlation functional; 
same Gaussian basis sets are used for all the calculations. Our results are summarized in Fig. 5. Though all different functionals 
(SVWN, PW91PW91, PW91LYP)\cite{Gaussian} yield similar current-voltage characteristic, the magnitude of the current is found to be much higher 
($\sim$ 3 times at a $V_{sd}$ of $\sim$0.7 V) than that obtained with the B3LYP approach. An atomic self interaction corrected 
DFT scheme yielding a lower conductance than the conventional DFT approach has been reported in a molecular junction\cite{Sanvito}. The inclusion of part of the exact exchange 
from the Hartree-Fock formalism in our {\it posteriori} B3LYP approach corrects partly the self-interaction error that occurs in the 
conventional density functional method; it represents a substantial improvement in the right direction as evident from the $V_{sd}-I_{sd}$ curve (Fig. 5).  

\subsection{ Bias Dependent Transmission}

 To investigate the intriguing features in  gate field induced current and to understand the origin of the field effect behavior in PbSNW, we have 
calculated the bias dependent transmission function as a function of injection energy($E$) for different $V_g$(Fig. 6). For brevity, we have only considered 
 $V_{sd}$  $\sim$0.76 V. First, the increase of area under the transmission curve within the chemical potential window (CPW) with the increase of 
negative gate bias confirms the observed increase of $I_{sd}$ with $V_g$ (Fig. 3); 
the non-linear increase in area explains the change of slope in $I_{sd}$-$V_g$ plot presented in the inset of Fig. 3. 
Analysis of eigenvalues of Hamiltonian for the NW reveals unoccupied levels (shown in Fig. 6) contributes to the conduction. Increasing the $V_g$, 
the participating unoccupied eigen-channel shifts in the direction of Fermi energy. For $V_g$=0 V, only L0 
level contributes to the $T(E,V)$ within the CPW. As $V_g$ increases more unoccupied levels move into the CPW, resulting in an increase in the 
density of states within the CPW. To quantify the response of the gate field, we have plotted the Stark shift ( $\epsilon_{g}^{i}-\epsilon_{0}^{i}$;
 $i$-corresponds to different unoccupied levels, $\epsilon_{g}$ and  $\epsilon_{0}$ are respectively the orbital energy in the presence and absence of 
gate field) as a function of $V_g$ for different participating unoccupied levels in Fig. 7. 
A significant Stark shift  has been observed. Different levels exhibit different shift, particularly at higher $V_g$. A closer examination indicates 
a non linear increase of Stark shift ($\Sigma_{i}\alpha_i \varepsilon_i + \frac{1}{2}\Sigma_{i,j}\beta_{ij} \varepsilon_i \varepsilon_j +\ldots $ )
 with the increase of $V_g$. 

\begin{figure}
\epsfig{figure=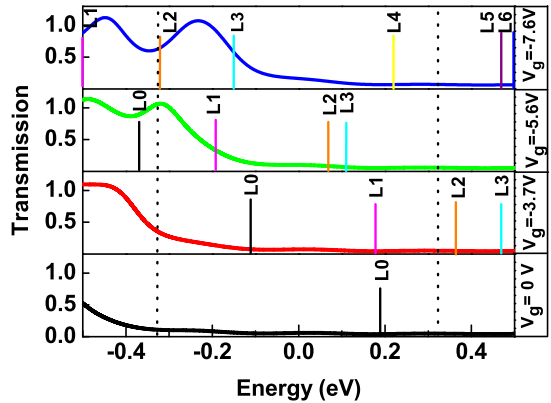,width=230pt}
\caption{ (Color online) Bias dependent transmission function as a function of injection energy for different gate bias at $V_{sd}$ $\sim$ 0.6 V.
The Fermi energy is set to zero in the energy scale; dotted lines represent the chemical potential window.
Notation: L0, L1, L2, L3, and L4 refer to LUMO, LUMO+1, LUMO+2, LUMO+3, and LUMO+4.
 }
\end{figure}

\begin{figure}
\epsfig{figure=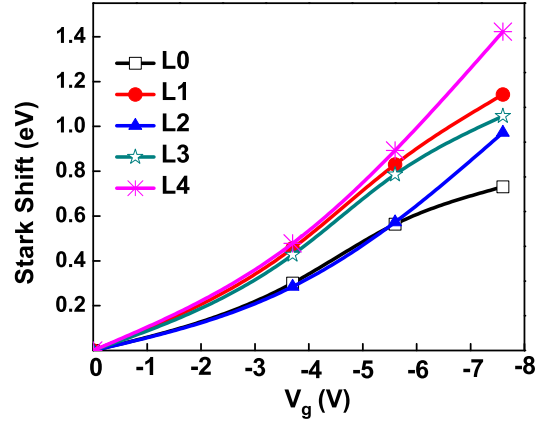,width=230pt}
\caption{(Color online)
Stark shift as a function of gate bias. Notation: L0, L1, L2, L3, and L4 refer to LUMO, LUMO+1, LUMO+2, LUMO+3, and LUMO+4.
A fixed $V_{sd}$ of $\sim$ 0.6 V is used. }
\end{figure}

\begin{figure}
\epsfig{figure=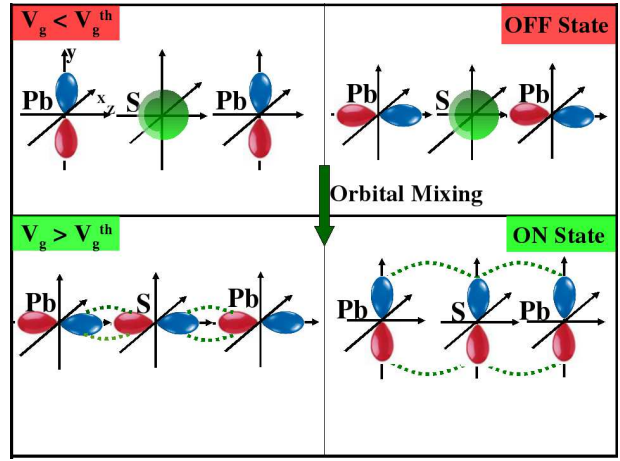,width=230pt}
\caption{ (Color online)
Schematic representation of orbital controlled mechanism for the PbSNW FET.
In the ON state ($V_g > V_{g}^{th}$), orbital mixing produces {\it p}-component at the S atoms resulting in an {\it inter-layer} orbital interaction along the {\it channel}
(z) axis.
The S-atom has only {\it s}-component in the OFF state ($V_g < V_{g}^{th}$).
}
\end{figure}

\subsection{Orbital Analysis}

Next, the natural question to ask is: How does the participating orbitals evolve with the gate bias? Does it have any correlation with the observed
increase in transmission in Fig. 6? To answer these subtle questions, we have analyzed participating MO coefficients in the presence and absence 
of gate field. As expected, for $V_g$=0V, the MOs are symmetric along the direction perpendicular to the wire axis (negative Y axis). Increasing the gate bias to -7.6 V,
the symmetry of the wavefunction breaks along the negative Y axis; the participating MOs localize
 in the same direction resulting the observed Stark shift
 (Fig. 7). A close inspection of the MO coefficients reveals that S atoms in the participating MO have only $s$-components in the absence of gate bias.
Increasing the gate bias beyond the threshold value of -3.7 V, $p$-components develop at the S atoms due to the strong gate field induced orbital mixing 
(Fig. 8). It should be noted that the Pb atom, which has $p$-component prior to the application of gate field, does not exhibit such orbital evolution. The 
$p$-components at the S atoms for the higher gate bias mediates {\it inter-layer} orbital interaction - allowing electron to delocalize along the
channel axis (Fig. 8). This explains unambiguously the origin of switching in conductance value observed in Fig. 3.
It is worth mentioning that very recently orbital gating has been observed in molecular junctions\cite{Reed2}.

\section{Conclusions}

  We present a new orbital-control mechanism to explain the gate field induced switching of current in a semiconducting PbSNW junction. 
An implicit orbital dependent single particle Green's function approach that employes a self-interaction correction scheme is used to calculate 
the electronic current. A comparative study using different exchange correlation functionals shows a quantitative improvement in the magnitude of current
for the self-interaction corrected scheme over the conventional DFT. 
 Both first and higher order Stark
effects are included in our model. 
The consistent increase of calculated current upon increasing negative gate bias as observed in the experiment, and 
the similar orbital evolution in a PbSe nanowire junction upon application of gate field reassure the validity of our generalized model, 
which can also be used to understand 
switching of current in other lead-chalcogenide NW junctions. Thus, the present work may serve as a guiding point in designing orbital-controlled 
nanowire-FET for potential applications in new generation electronic circuit.

\section{Acknowledgement}
This work is supported by NSF through Grant No. 0643420.


\section{ References}




\begin{thebibliography}{99}

\bibitem{lieber1}
J. Xiang, W. Lu, Y. Hu, Y. Wu, H. Yan, and C. M. Lieber, Nature {\bf 441}, 489 (2006).
\bibitem{circuit}
H. Yan, H. S. Choe, S. Nam, Y. Hu, S. Das, J. F. Klemic, J. C. Ellenbogen, and C. M. Lieber, Nature {\bf 470}, 240 (2011).
\bibitem{ScienceT}
D. V. Talapin and C. B. Murray, Science {\bf 310}, 86 (2005).
\bibitem{Talapin1}
K-S. Cho, D. V. Talapin, W. Gaschler, and C. B. Murray, J. Am. Chem. Soc. {\bf 127}, 7140 (2005).
\bibitem{Talapin2}
 D. V. Talapin, C. T. Black, C. R. Kagan, E. V. Shevchenko, A. Afzali, C. M. Murray, J. Phys. Chem. C {\bf 111}, 13244 (2007).
\bibitem{Talapin3}
 J-S. Lee, E. V. Shevchenko, and D. V. Talapin, J. Am. Chem. Soc. {\bf 130}, 9673 (2008).
\bibitem{pbsnw1}
I. Patla, S. Acharya, L. Zeiri, J. Israelachvili, S. Efrima, Y. Golan, Nano Lett.
{\bf 7} 1459 (2007).
\bibitem{pbsnw2}
P. K. Mukherjee, K. Chatterjee, D. Chakravorty.  Phys. Rev. B {\bf 73} 035414 (2006).
\bibitem{pbsnw3}
F. Gao, Q. Lu, X. Liu, Y. Yan, D. Zhao, Nano Lett. {\bf 1} 743 (2001).
\bibitem{ACS}
S. Y. Jang, Y. M. Song, H. S. Kim, Y. J. Cho, Y. S. Seo, G. B. Jung, C.-W. Lee, J. Park, M. Yung, J. Kim, B. Kim, J-G. Kim, Y.-J. Kim,
ACS NANO {\bf 4}, 2391 (2010).
\bibitem{Max}
M. Di Ventra, {\it Electrical Transport in Nanoscale Systems}, (Cambridge, New York, 2008).
\bibitem{dutta}
S. Datta, {\it Electron Transport in Mesoscopic Systems} (Cambridge University Press, Cambridge, England, 1997).
\bibitem{Ratner}
A. Nitzan and M. A. Ratner, Science {\bf 300}, 1384 (2003).
\bibitem{Guo1}
D. Waldron, P. Haney, B. Larade, A. MacDonald, and H. Guo, Phys. Rev. Lett. {\bf 96}, 166804 (2006).
\bibitem{Guo2}
J. Taylor, H. Guo, J. Wang, Phys. Rev. B {\bf 63}, 245407 (2001).
\bibitem{Sanvito}
C. Toher, and S. Sanvito, Phys. Rev. Lett. {\bf 99}, 056801 (2007).
\bibitem{Pati}
R. Pati, M. McClain, A. Bandyopadhyay, Phys. Rev. Lett. {\bf 100}, 246801 (2008).
\bibitem{Gaussian}
Gaussian 03, Gaussian Inc., Pittsburgh, PA, 2003.
\bibitem{Subhasish}
S. Mandal, and R. Pati, Chem. Phys. Lett. {\bf 479}, 312 (2009).
\bibitem{Subhasish2}
S. Mandal, and R. Pati, Phys. Rev. B {\bf 83}, 195420 (2011). 
\bibitem{PPP}
P. P. Pal, and R. Pati, Phys. Rev. B {\bf 82}, 045424 (2010).
\bibitem{evers}
F. Evers, F. Weigend, M. Koentopp, Phys. Rev. B {\bf 69}, 235411 (2004).
\bibitem{Runge}
E. Runge and E. K. U. Gross, Phys. Rev. Lett. {\bf 52}, 997 (1984).
\bibitem{Max2}
N. Sai, M. Zwolak, G. Vignale, M. Di Ventra, Phys. Rev. Lett. {\bf 94}, 186810 (2005).
\bibitem{Max3}
G. Vignale and M. Di Ventra, Phys. Rev. B {\bf 79}, 014201 (2009).
\bibitem{Reed2}
H. Song, Y. Kim, Y. H. Jang, H. Jeong, M. A. Reed, T. Lee, Nature {\bf 462}, 1039 (2009).



\end{thebibliography}
\end{document}